\documentstyle[amssymb,aps,prl,twocolumn]{revtex}

\begin{document}

\twocolumn[\hsize\textwidth\columnwidth\hsize\csname
@twocolumnfalse\endcsname
\draft
\title{Universal scaling of Hall resistivity in clean and moderately clean limits
for Hg- and Tl-based superconductors}
\author{W. N. Kang,$^{1}$ Wan-Seon Kim,$^{1}$ S. J. Oh,$^{1}$ Sung-Ik Lee,$^{1}$ D.
H. Kim,$^{2}$ C. H. Choi,$^{3}$ H. -C. Ri,$^{3}$ and C. W. Chu$^{4}$}
\address{$^{1}$ National Creative Research Initiative Center for Superconductivity,
Department of Physics, Pohang University of Science and Technology, Pohang
790-784, Korea}
\address{$^{2}$ Department of Physics, Yeungnam University, Kyungsan 712-749, Korea}
\address{$^{3}$ Material Science Team, Joint Research Division, Korea Basic Science
Institute, Taejon 305-333, Korea}
\address{$^{4}$ Texas Center for Superconductivity, University of Houston, Houston,
TX 77204, USA}
\maketitle

\begin{abstract}
The mixed-state Hall resistivity $\rho _{xy}$ and the longitudinal
resistivity $\rho _{xx}$ in HgBa$_{2}$CaCu$_{2}$O$_{6}$, HgBa$_{2}$Ca$_{2}$Cu$_{3}$O$_{8}$, and Tl$_{2}$Ba$_{2}$CaCu$_{2}$O$_{8}$ thin films have been
investigated as functions of the magnetic field (H) up to 18 T. We observe the
universal scaling behavior between $\rho _{xy}$ and $\rho _{xx}$ in the
regions of the clean and the moderately clean limit. The scaling exponent $\beta $ in $\rho _{xy}=A\rho _{xx}^{\beta }$ is $1.9\pm 0.1$ in the clean
limit at high H and low temperature (T) whereas $\beta $ is $1\pm 0.1$ in
the moderately clean limit at low H and high T, consistent
with a theory based on the midgap states in the vortex core. This finding
implies that the Hall conductivity $\sigma _{xy}$ is also universal in Hg- and Tl-based
superconductors.

\end{abstract}


\vskip 0.5pc]

\newpage 
When a type II superconductor is cooled down from a normal state
into a superconducting state, the Hall effect shows very unusual features,
which have been a long-standing problem and have remained as unresolved
issues for more than three decades. The sign reversal of the Hall effect
below T$_c$ is one of the most interesting phenomena in the flux dynamics 
for high-T$_{c}$ superconductors
(HTS) and has attracted both experimental and theoretical interest.
Furthermore, a scaling behavior between $\rho _{xy}$ and $\rho _{xx}$ has
been found in most HTS \cite
{Luo92,Samoilov93,Budhani93,Kang96,Kang97,Kang99a}. The puzzling scaling
relation, $\rho _{xy}=A\rho _{xx}^{\beta }$, with $\beta \sim 2$ has been
observed for Bi$_{2}$Sr$_{2}$CaCu$_{2}$O$_{8}$ (Bi-2212) crystals \cite
{Samoilov93} and Tl$_{2}$Ba$_{2}$Ca$_{2}$Cu$_{3}$O$_{10}$ (Tl-2223) films 
\cite{Budhani93}. Other similar studies have found $\beta =1.5\sim 2.0$ for
YBa$_{2}$Cu$_{3}$O$_{7}$ (YBCO) films \cite{Luo92}, YBCO crystals \cite
{Kang96}, and HgBa$_{2}$CaCu$_{2}$O$_{6}$ (Hg-1212) films \cite{Kang97}.
Even more interestingly, $\beta \sim 1$ was observed in the Hg-1212 thin
films \cite{Kang99a} after heavy-ion irradiations.

To interpret this scaling behavior, a number of theories have been proposed.
The first theoretical attempt was presented by Dorsey and Fisher \cite
{Dorsey92}. They showed that near the vortex- grass transition, $\rho _{xy}$
and $\rho _{xx}$ could be scaled with an exponent $\beta =1.7$, and they
explained the experimental results of Luo {\it et al.} for YBCO films \cite
{Luo92}. A phenomenological model was put forward by Vinokur {\it et al.} 
\cite{Vinokur93}. They claimed that in the thermally assisted flux-flow
(TAFF) region, $\beta $ should be 2 and independent of the pinning strength.
Their result was consistent with the observed exponent in Bi-2212 crystals
and Tl-2223 films only for high H. Another phenomenological model was
proposed by Wang {\it et al.} \cite{Wang94}. They showed that $\beta $ could
change from 2 to 1.5 as the pinning strength increased, which agreed with
the results reported for YBCO crystals \cite{Kang96} and Hg-1212 films \cite
{Kang97}. However, all these theories fail to explain the wide range of $%
\beta $ from 1 to 2 observed in ion-irradiated Hg-1212 films \cite{Kang99a}.

Recently, a more detailed theory based on localized states in vortex cores
was developed by Kopnin and Lopatin (KL) \cite{Kopnin95} for the clean limit
(CL) and the moderately clean limit (MCL). Due to the short coherence
lengths, HTS change from the MCL to the CL as T decreases from
T$_{c}$. KL showed that $\sigma _{xy}=\rho _{xy}/\rho _{xx}^{2}$ was
universal in the CL whereas the tangent of the Hall angle was universal in
the MCL. This implies that $\beta $ can change from 1 to 2 with decreasing
T, which is consistent with previous work on ion-irradiated
Hg-1212 \cite{Kang99a}. This theory also well describes the recent
observation of a triple Hall sign reversal in Hg-1212 thin films containing
a high density of columnar defects \cite{Kang99b}. Localized core states
have been observed in HTS by using various experimental setups, such as
far-infrared spectroscopy \cite{Karrai92} and scanning tunneling
spectroscopy \cite{Maggio95}, and they are in good agreement with the
theoretical predictions \cite{Morita97}. Furthermore, in the MCL case,
Kopnin and Volovik \cite{Kopnin98} showed that $\sigma _{xy}$ for d-wave
superconductors was very similar to the result \cite{Kopnin95} for s-wave
superconductors. On the other hand, Frantz and Tesanovic \cite{Franz98}
claimed that a bound state does not exist in the vortex core of a d-wave
superconductor.

In this Letter, we report the first demonstration of the universal scaling
behavior of the Hall resistivity in the CL and the MCL regions for Tl- and
Hg-Based Superconductors, and the results can be well described by the
recent KL theory. In the present study, by using a low-noise preamplifier
prior to a nanovoltmeter, we were able to expand the sensitivity of the Hall
voltage up to one order of magnitude compared to the sensitivities in
previous works, and we were able to confirm the universality of the Hall 
scaling behavior for an extended H range up to 18 T.

The transport properties and fabrication process of Hg-based superconducting
thin films are described in detail elsewhere \cite{Kang98,Kang99c}. The
Tl-2212 thin films are commercially available \cite{STI}. The typical
dimensions of the thin films were 5 $mm\times $10 $mm\times 0.5-$1 $\mu m$.
The mid-resistance T, T$_{c}$, in zero H for the Hg-1223, the
Hg-1212, and the Tl-2212 films were 132, 127, 106 K, respectively. The X-ray
diffraction patterns indicated highly oriented thin films with the c axes
normal to the plane of the substrate and phase purities of more than 95 $\%$%
. The transition width was found to be less than 2 K. The heavy-ion irradiation of the
Tl-2212 films was performed along their c axes by using 1.4-GeV U ions. The
irradiation dose was 6 $\times $ 10$^{10}$ ions/cm$^{2}$, which corresponded
to a matching H, B$_{\phi }$, of ~ $\thicksim $ 1.2 T. The Hg-1212 films
were irradiated at a dose of 5 $\times $ 10$^{10}$ ions/cm$^{2}$ (B$_{\phi }$
$\thicksim $ 1 T) along the c axes by using 5-GeV Xe ions \cite{Kang99a}.
The values of $\rho _{xy}$ and $\rho _{xx}$ were simultaneously measured
using a two-channel nanovoltmeter (HP34420A) and the standard five-probe dc
method. A low-noise preamplifier (N11, EM Electronics Inc.) was installed
prior to the nanovoltmeter in order to increase the sensitivity of the Hall
voltage at low H. The applied dc current density was 100 $-$ 250 A/cm$%
^{2}$. Both $\rho _{xx}$ and $\rho _{xy}$ were Ohmic at these current levels.
H was applied parallel to the c axes of the thin films. The value of $\rho _{xy}$
was extracted from the antisymmetric part of the Hall voltages measured under opposite H.

Figure 1 shows T dependences of $\rho _{xx}$ and $\rho _{xy}$
for Hg-1212 and Hg-1223 thin films for various H up to 18 T.
In order to compare $\rho _{xx}$ and $\rho _{xy}$ for different samples, we
use a reduced-T scale, T/T$_{c}$, rather than a real-T
scale. At low H, the difference in $\rho _{xx}$ between Hg-1212 and
Hg-1223 is clearly visible and decreases with increasing H up to 18 T,
showing that the addition of one CuO$_{2}$ layer increases T$_{c}$ but
weakens the pinning strength at low H. Interestingly, the behavior of $%
\rho _{xy}$ is different from that of $\rho _{xx}$. A significant difference
in $\rho _{xy}$ can be observed even at 18 T, suggesting that the flux-flow $%
\rho _{xy}$ is more sensitive than the flux-flow $\rho _{xx}$ to the number
of CuO$_{2}$ layers. A comparison of the physical properties for those two
samples may provide an interesting explanation for the role of the CuO$_{2}$
layers \cite{Lokshin98} in the homologous series of HgBa$_{2}$Ca$_{n-1}$Cu$%
_{n}$O$_{2n+2}$ superconductors with n = 1 $-$ 6. The $\rho _{xx}$ and $\rho
_{xy}$ data for heavy-ion irradiated Hg-1212 are shown in Ref. 6.

In Fig. 2, we show T dependences of $\rho _{xx}$ and $\rho
_{xy}$ for Tl-2212 thin films before and after irradiations. Although T$_{c}$
at zero H decreases by $\thicksim $ 2 K after irradiation, the large
enhancement of T$_{c}$ in H due to the strong pinning by
columnar defects is clearly observed. This agrees with previous observations 
\cite{Budhani93} for irradiated samples. The scaling behaviors between $\rho
_{xy}$ and $\rho _{xx}$ for Hg-1212 and Hg-1223 films for various H up to 18 T are plotted in Fig. 3. The corresponding data for Tl-2212
films before and after irradiations are shown in Fig. 4. Since $\rho _{xy}$
below H = 2 T is negative in a certain T region, we plot the
absolute value $|\rho _{xy}|$. The $\beta $ in $|\rho _{xy}|=A\rho
_{xx}^{\beta }$ is extracted from the slope of the solid lines, as shown in
Figs. 3 and 4. Hall scaling is observed over roughly two decades of $\rho
_{xy}$, and even four decades in high H. This scaling relation is valid
in the TAFF region. Note that the TAFF region expands to lower T
at high H due to a huge resistive broadening in H for
these materials. Thus, we can investigate the Hall behavior in the clean
limit by applying high H \cite{Harris94}. On the other hand,
the low-H data correspond to the MCL since the TAFF
region in this case is limited to near T$_{c}$.

H dependence of the Hall scaling is clearly demonstrated by the
above data, and the results, including previously observed data for
irradiated Hg-1212 films, are summarized in Fig. 5. As H increases, $%
\beta $ changes from 1.4 to 1.9 for Hg-1212, from 1.3 to 1.9 for Hg-1223,
from 1.0 to 1.9 for the pristine Tl-2212, from 1.0 to 1.9 for the irradiated
Tl-2212, and from 1.0 to 1.9 for the irradiated Hg-1212 \cite{Kang99a}. Note
that at higher H, $H\geq 8$ T, the scaling exponent $\beta =1.9\pm 0.1$
shows a universal behavior, regardless of H, the number of CuO$_{2}$
layers, the types of defects, and even the types of compounds. More
strikingly, at low H, $\beta =1\pm 0.1$ also appears as a universal
number although the observed H range is rather limited. The scaling
exponent is independent of H below H = 0. 3 T for pristine Tl-2212 and
below H = 1.2 T for irradiated Tl-1212 and Hg-1212 films. This universal
behavior of the scaling is our principal finding, and this observation has
serious implications for the physics of the Hall behavior, as discussed
below.

With short coherence lengths and large energy gaps in HTS, the discrete
nature of the energy levels, $\omega _{o}$, in the vortex cores has been
observed experimentally \cite{Karrai92,Maggio95} and has been interpreted
theoretically \cite{Morita97}. Considering these localized states and an
additional force induced by the kinetic effects of charge imbalance
relaxation, KL \cite{Kopnin95} calculated the Hall and the longitudinal
conductivities, $\sigma _{L}$, in the CL and the MCL regions. According to this theory, the
Hall conductivity can be described by three terms: $\sigma
_{H}= $ $\sigma _{H}^{(L)}+\sigma _{H}^{(D)}+\sigma _{H}^{(A)}$, where $%
\sigma _{H}^{(L)}$, $\sigma _{H}^{(D)}$, and $\sigma _{H}^{(A)}$ are the
contributions from localized excitations, delocalized excitations, and an
additional force, respectively. The additional force is determined by the
energy derivative of the density of states at the Fermi surface. Since $%
\sigma _{H}^{(A)}$ dominates over $\sigma _{H}^{(L)}$ and $\sigma _{H}^{(D)}$%
\ near T$_{c}$, the Hall anomaly can take place, as observed in most HTS. $%
\sigma _{H}^{(D)}$ originates from the density of quasiparticles outside the
vortex core; thus, $\sigma _{H}^{(D)}$ is comparable to the normal-state
Hall conductivity very near T$_{c}$, but is very small at low T
compared to $\sigma _{H}^{(L)}$ . Due to this, we can neglect $\sigma
_{H}^{(A)}$ and $\sigma _{H}^{(D)}$\ in the low-T region. Note
that the Hall scaling behavior is observed in the TAFF regions, which
correspond to T regions below the positive peaks in the $\rho
_{xy} $ $-$ T curves. In the TAFF regions, therefore, $\sigma _{H}$ and 
$\sigma _{L}$, can be expressed by \cite{Kopnin95} 
\begin{equation}
\sigma _{H}\sim \frac{Ne}{B}\frac{(\omega _{o}\tau )^{2}}{1+(\omega _{o}\tau
)^{2}},
\end{equation}
\begin{equation}
\sigma _{L}\sim \frac{Ne}{B}\frac{\omega _{o}\tau }{1+(\omega _{o}\tau )^{2}}%
,
\end{equation}
where N is the density of charge carriers and $\tau $ is the relaxation
time. It has been found \cite{Kopnin95,Harris94} that the tangent of the
Hall angle, $tan\Theta =\sigma _{H}/\sigma _{L}\thicksim \omega _{o}\tau $,
is very small ($\ll $ 1) in the dirty region near T$_{c}$ while it is very
large ($\gg $ 1) in the superclean region T $\ll $ T$_{c}$. According to
Eqs. (1) and (2), there are two distinct scaling regions. For the
low-T (CL) region with $(\omega _{o}\tau )^{2}\gg 1$, Eq. 1
becomes $\rho _{xy}=(Ne/B)\rho _{xx}^{2}$, resulting in a universal scaling
law of $\beta =2$, which is also predicted by the phenomenological models
proposed by Vinokur {\it et al.} \cite{Vinokur93} and Wang {\it et al.}\cite
{Wang94}. However, at relatively high T (MCL) with $(\omega
_{o}\tau )^{2}\ll 1$, we obtain $\rho _{xy}=(\omega _{o}\tau)\rho _{xx}$; thus, a
universal scaling law with $\beta =1\pm 0.1$ should be observed since $\rho
_{xx}$ is an exponential function of T while $\omega _{o}\tau $ is
a slowly varying function of T \cite{Harris94}. These features are
explicitly consistent with our present results shown in Fig. 5. In other
words, a universal values of $\beta =2$ and $\beta =1$ are found for the CL
and the MCL, respectively. In the crossover regions from the MCL to the CL, $%
1<\beta <2$ is found. These observations were possible because of the large
resistive broadening in H for Hg- and Tl-based compounds
and were partially due to the enhanced sensitivity of the measurement. For
the irradiated samples, the universal regions shifted to higher H,
indicating that the MCL and the CL regions moved to lower T due to
the impurity effect of the columnar defects.

In the case of YBCO, however, the Hall scaling \cite{Luo92,Kang96} could be
different from those observed for Hg- and Tl-based superconductors. Since
the Hall scaling in YBCO takes place for $\rho _{xy}<0$, where $\sigma
_{H}^{(A)}$ is comparable to $\sigma _{H}^{(L)}$, the Hall scaling can be
modified by T dependence of $\sigma _{H}^{(A)}$. Furthermore,
because $\sigma _{H}^{(A)}$ is more pronounced with increasing T,
the scaling range of $\rho _{xy}$ is narrower than those observed in Tl- and
Hg-based superconductors. This is a possible explanation why $\beta =1$ has
not been observed in YBCO.

In summary, the universal Hall scaling behaviors between $\rho _{xy}$ and $%
\rho _{xx}$ in Hg-1212, Hg-1223, and Tl-2212 thin films were investigated as
functions of H. We found the universal behavior of the
Hall scaling for the CL and the MCL regions. Within the context of a recent
theory \cite{Kopnin95} based on the localized states in vortex cores, this
universal behavior was explicitly understood. However, this behavior is
valid only if $\sigma _{H}^{(L)}$ is the main contribution to the Hall
effect, which is not the case for YBCO.

\bigskip

This work is supported by the Creative Research Initiatives of the Korean
Ministry of Science and Technology.

\begin{figure}[tbp]
\caption{Reduced-T dependences of $\protect\rho _{xx}$ (top) and $%
\protect\rho _{xy}$  (bottom) curves for Hg-1212 (solid lines) and for
Hg-1223 (open circles) thin films.}
\end{figure}
\begin{figure}[tbp]
\caption{T dependences of $\protect\rho _{xx}$ (top) and $\protect%
\rho _{xy}$  (bottom) curves for Tl-2212 thin films before (solid lines) and
after (open circles) ion irradiations.}
\end{figure}
\begin{figure}[tbp]
\caption{Scaling behaviors between $\protect\rho _{xy}$ and $\protect\rho %
_{xx}$ for Hg-1212 (top) and Hg-1223 (bottom) thin films.}
\end{figure}
\begin{figure}[tbp]
\caption{Scaling behaviors between $\protect\rho _{xy}$ and $\protect\rho %
_{xx}$ for Tl-2212 thin films before (solid lines) and after (open circles)
ion irradiations. $\protect\beta =1.0$ is observed for pristine films at H =
0.3 T and for irradiated films for $H<B_{\protect\phi }$.}
\end{figure}
\begin{figure}[tbp]
\caption{H dependences of the scaling exponent $\protect\beta $ for
various Hg- and Tl-based HTS before (open symbols) and before (solid
symbols) ion irradiations. The fact that the universal power law has $%
\protect\beta =1.9$ above H = 8 T and $\protect\beta =1.0$ at low H,
which are independent of H, is clearly visible.}
\end{figure}

\end{document}